\documentclass[usenatbib]{mnras}
\usepackage{graphicx}	
\usepackage{multicol}	
\usepackage[english]{babel}
\usepackage{epstopdf}
\usepackage[T1]{fontenc}
\usepackage{ae,aecompl}
\usepackage{amsmath}    
\usepackage{amssymb}    
\usepackage{ulem}

\def\mpc{\,h^{-1}{\rm Mpc}}

\def\msun{\,h^{-1}{\rm M}_\odot}

\def\bx{{\boldsymbol x}}

\def\knot{{\tt knot}}
\def\sheet{{\tt sheet}}
\def\filament{{\tt filament}}
\def\void{{\tt void}}

\usepackage{color}

\makeatletter

\newcommand{\Rmnum}[1]{\expandafter\@slowromancap\romannumeral #1@}
\makeatother

\title[galaxy-halo connection in the cosmic web]{Connections between galaxy properties and halo formation
time in the cosmic web}

\author[Youcai Zhang]{Youcai Zhang$^{1}$\thanks{yczhang@shao.ac.cn},
Xiaohu Yang$^{2,3}$,
Hong Guo$^{1}$
\\
$^{1}${Key Laboratory for Research in Galaxies and Cosmology,
  Shanghai Astronomical Observatory; Nandan Road 80, Shanghai 200030,
  China} \\
$^{2}$Department
of Astronomy, School of Physics and Astronomy, and Shanghai Key Laboratory for Particle Physics and Cosmology, \\
~~Shanghai Jiao Tong University, Shanghai 200240, China \\
$^{3}$Tsung-Dao Lee Institute and Key Laboratory for
    Particle Physics, Astrophysics and Cosmology, Ministry of Education, \\
    ~~Shanghai Jiao Tong University, Shanghai 200240, China
}

\begin{document}
\label{firstpage}
\pagerange{\pageref{firstpage}--\pageref{lastpage}}
\maketitle

\begin{abstract}
By linking galaxies in Sloan Digital Sky Survey (SDSS) to subhaloes in the ELUCID simulation, we investigate the relation between subhalo formation time and the galaxy properties, and 
the dependence of galaxy properties on the cosmic web environment. We find that central and satellite subhaloes have different formation time, where satellite subhaloes are older than central subhaloes at fixed mass. At fixed mass, the galaxy stellar-to-subhalo mass ratio is a good proxy of the subhalo formation time, and increases with the subhalo formation redshifts, especially for massive galaxies. The subhalo formation time is dependent on the cosmic web environment. For central subhaloes, there is a characteristic subhalo mass of $\sim 10^{12} \msun$, below which subhaloes in knots are older than subhaloes of the same mass in filaments, sheets, or voids, while above which it reverses. The cosmic web environmental dependence of stellar-to-subhalo mass ratio is similar to that of the subhalo formation time.  For centrals, there is a characteristic subhalo mass of $\sim 10^{12} \msun$, below which the stellar-to-subhalo mass ratio is higher in knots than in filaments, sheets and voids, above which it reverses. Galaxies in knots have redder colors below $10^{12} \msun$, while above $10^{12} \msun$, the environmental dependence vanishes. Satellite fraction is strongly dependent on the cosmic web environment, and decreases from knots to filaments to sheets to voids, especially for low-mass galaxies.
\end{abstract}


\section{Introduction}\label{sec_intro}

In the current standard $\Lambda$CDM model, galaxies are believed to form through baryonic
gas cooling and condensation in dark matter haloes which reside in the context of cosmic web
composed of knots, filaments, sheets and voids \citep{White1978,Bond1996}. Therefore, the 
properties of galaxies are expected to closely linked to their host haloes, as well as the 
large-scale environment \citep{Lim2016,Correa2020,Xu2020}. The physical connection among galaxies, haloes and the
cosmic web environment is an interesting and challenging problem in the study of
galaxy formation and evolution \citep{Musso2018,Lange2019}.

From cosmological $N$-body simulations, it's well established that the clustering of haloes 
depends not only on the halo mass, but also other properties of dark matter haloes, such as the halo formation 
time. This is so called {\tt halo assembly bias} \citep{Sheth2004, Gao2005, Wechsler2006,
Gao2007,Zentner2014,Contreras2019, Mansfield2020}. Using $N$-body simulation, \citet{Gao2005} 
showed that the clustering of haloes are strongly dependent on the
halo formation time. For haloes at low-mass end, older haloes cluster more strongly than 
younger haloes of the same mass. For haloes at high-mass end, \citet{Jing2007} found that
the age dependence is reversed, where older haloes are less clustered than younger haloes of
the same mass. The dependence of halo clustering on halo formation time besides
mass is also extended to other halo properties, such as concentration, spin, shape 
\citep{Wechsler2006,Lacerna2012,Villarreal2017}.

In observation, there is no consensus on a robust determination of the assembly bias
in the galaxy distribution, referred to as {\tt galaxy assembly bias}. Although
some studies claimed that the assembly bias in galaxy distribution is significantly robust
\citep{Yang2006,Cooper2010,WangL2013,Lacerna2014,Miyatake2016,Ferreras2019,Obuljen2020,Yuan2020}, 
others argued that the galaxy assembly bias signal is 
small or negligible and may be a result of the systematic uncertainties 
\citep{Blanton2007b, Tinker2008,Lin2016, Zu2016, Dvornik2017, Zu2017,Tinker2017, Tinker2018, Salcedo2020}. 
Assuming that redder galaxies live in older haloes, \citet{Hearin2013a}
developed an age-matching model by extending the traditional abundance-matching technique.
In the age-matching model, the galaxy colour at fixed luminosity is assumed to be
monotonically correlated with the halo age at fixed $V_{\rm max}$, which is the halo 
maximum circular velocity.   \citet{Hearin2014} claimed that the age-matching model 
can accurately predict colour-dependent galaxy clustering  and a variety of galaxy group statistics.
In contrast, \citet{Zu2016} found that the age-matching model fails to reproduce the observed halo
mass for massive blue centrals, using the galaxy-galaxy lensing measurements split by colour in
bins of galaxy stellar mass. \citet{Tinker2017} claimed that the measurements of the quenched 
fraction of centrals can not be reconciled with the prediction of the age-matching model for 
low-mass galaxies, implying that there is no relation between halo formation time and galaxy
quenching for centrals. 

In some previous studies \citep{Yang2006, WangL2013, Hearin2013a,Hearin2014,Watson2015}, one key assumption is
that the galaxy color or specific star formation rate is a good indicator of halo formation time.
However, \citet{Lin2016} found no convincing evidence of galaxy assembly bias using specific
star formation rate as a proxy of halo formation time, which indicates that the galaxy sSFR 
is not well correlated with the halo formation history. Because of the inconsistent studies of galaxy assembly bias in observation, one of the purposes in this work is to find which galaxy property is tightly correlated with the halo formation time.

In this work, with the help of the constrained ELUCID simulation, the properties of {\tt ture} galaxies 
in observation can be related with the formation time of dark matter subhaloes in simulation, 
so that we can directly check which property is the best proxy of halo formation time. Using the ELUCID
simulation, galaxies in observation can be linked to dark matter subhaloes in simulation using a 
neighborhood abundance matching method \citep{Yang2018}, since the mass and positions of haloes at $z \sim 0$
in ELUCID simulation are consistent with those of galaxy groups in observation, due to the initial condition 
of the ELUCID simulation constrained from the density field of galaxy distribution in observation
\citep{WangHuiyuan2012, WangHuiyuan2014, WangHuiyuan2016, Zhang2021}.

Besides, in order to interpret galaxy or halo assembly bias, recent studies have focused on the key role
of the cosmic web environment \citep{Tojeiro2017, Yang2017, Musso2018, Sinigaglia2020}, 
and claimed that the tidal anisotropy of cosmic web environment
is the primary indicator of halo assembly bias \citep{Paranjape2018a,Paranjape2018b,Ramak2019,XuXiaoju2020,Ramak2020}.
Therefore, to better understand the assembly bias, recent studies have focused on the influence of 
cosmic web environment on the galaxy or halo properties \citep{Hahn2007a,WangH2011,Chen2020,Xu2020}.

On the one hand, there are lots of evidence that the cosmic web environment affects galaxy
or halo properties \citep{Chen2017, Kraljic2018}. Using a series of $N$-body simulations, \citet{Hahn2007a} claimed that
for low-mass haloes, the formation redshifts strongly depend on the cosmic web environment, and
haloes of fixed mass tend to be older in knots and younger in voids. Using data from
the Galaxy And Mass Assembly (GAMA) survey, \citet{Alpaslan2016} found that galaxies closer to the cylindrical 
axes of the filaments have lower specific star formation rate at a given mass. Using galaxies
from SDSS DR 12, \citet{Chen2017} claimed that at a given stellar mass, older galaxies tend
to reside closer to the cylindrical axes of the filaments than younger galaxies. Similar results have 
also been obtained in observation or hydrodynamical simulation
\citep{Malavasi2017, Laigle2018, Mahajan2018, Luber2019, Singh2020}.

On the other hand, lots of studies supported that properties of galaxies or haloes are independent 
of their cosmic web location (geometric environment), but entirely determined by the environmental density 
\citep{Alonso2015,Alonso2016, Eardley2015,Brouwer2016, Goh2019}. In simulation, \citet{Alonso2015} 
showed that the halo mass function depends only on their local environmental density, and not on the 
cosmic web location. In observation, using data from GAMA survey, \citet{Eardley2015} also claimed that
the galaxy luminosity function is independent on the cosmic web location at the same environmental density.
Using the Bolshoi-Planck simulation, \citet{Goh2019} claimed that at the same environmental density,
there is no discernible difference between the halo properties (e.g. halo spin parameter, concentration, and
specific mass accretion rate) in different cosmic web environment of filaments, walls, and voids.
In addition, \citet{Alpaslan2015} claimed that the most important parameter driving galaxy
properties is the stellar mass, as opposed to the cosmic web environment.

Interestingly, using galaxy groups in GAMA survey, \citet{Tojeiro2017}
found that low-mass haloes show a steadily increasing galaxy stellar-to-halo mass ratio
from voids to knots, with the trend be reversed at large halo mass \citep[see also][]{Xu2020}.
This behavior indicates that at low-mass end, haloes in knots are older than haloes of the 
same mass in voids, if stellar-to-halo mass ratio is a good proxy of the halo formation time. Using galaxy groups from 
SDSS DR7, \citet{Lim2016} found that galaxy color is tightly correlated with stellar-to-halo
mass ratio for low-mass
central galaxies, and claimed that the ratio can be used as a proxy of the halo formation time. 

In this work, combining galaxies in SDSS DR7 and subhaloes in the ELUCID simulation, we mainly focus on the relation between subhalo formation time and the stellar-to-subhalo mass ratio. In addition, we investigate the dependence of galaxy properties and subhalo formation time on the cosmic web environment.

This paper is organized as follows. In Section~\ref{sec_data}, we describe the subhalo catalog
from the ELUCID simulation, and the observational data from SDSS DR7. In addition, we describe
the novel neighborhood subhalo abundance matching method linking galaxies in observation to
subhaloes in constrained simulation, and the cosmic web classification method. In Section~\ref{sec_proxy},
we investigate the relation between subhalo formation time and the galaxy properties.
In Section~\ref{sec_cw}, we study the dependence of galaxy properties on the cosmic web 
environment. Finally, we summarize our results in Section~\ref{sec_summary}. 
Throughout this paper, the cosmological parameters are $\Omega_m = 0.258$, $\Omega_b = 0.044$,
$\Omega_\Lambda = 0.742$, $h = 0.72$, $n_s = 0.963$ and $\sigma_8 = 0.796$.

\section{Data and methods}\label{sec_data}
\subsection{Subhaloes from ELUCID simulation}

The ELUCID simulation is a dark matter only, constrained simulation, which can reproduce the
density field of the nearby universe generated from the galaxy distribution in the SDSS 
observation. Therefore, the statistical properties of cosmic web are accurately reproduced 
in the ELUCID simulation, compared with the SDSS observation \citep{WangHuiyuan2016}.

The initial condition of the simulation is sampled with $3072^3$ dark matter particles at 
redshift $z = 100$ in a periodic box of $500 \mpc$ on a side. The density field is then 
evolved to the present epoch with a memory-optimized version of GADGET2 \citep{Springel2005b}
in the Center for High Performance Computing, Shanghai Jiao Tong University. In the 
simulation, $100$ snapshots are produced from redshift $z = 18.4$ to $z = 0$ with the 
expansion factor spaced in logarithmic space. The mass of each dark matter particle is 
$3.0875 \times 10^8 \msun$. The ELUCID simulation adopts the cosmological parameters:
$\Omega_m = 0.258$, $\Omega_b = 0.044$, $\Omega_\Lambda = 0.742$, $h = 0.72$, $n_s = 0.963$ and
$\sigma_8 = 0.796$.

From the simulation, the standard friends-of-friends (FOF) algorithm \citep{Davis1985} is 
performed on the particle data to generate FOF haloes with a linking length of $b = 0.2$ 
times the mean particle separation. Next, the SUBFIND algorithm \citep{Springel2001}  
decomposes a given FOF halo into a set of disjoint, gravitationally bound substructures 
in the unbinding procedure, in which the gravitational potentials are iteratively 
calculated to remove the unbound particles in the host FOF halo \citep{Springel2020}.
Within these substructures, the most massive one is called central subhalo, while the 
others are referred to satellite subhaloes. Compared with FOF haloes, a set of 
gravitationally bound subhaloes in simulation are more suitable to link central and 
satellite galaxies in observation. Therefore, in the following analysis, central 
subhaloes in simulation are linked to central galaxies in observation, while satellite 
subhaloes are matched to satellite galaxies \citep{Yang2018}.

For each central or satellite subhalo, the center of the subhalo is defined as the 
position of the most bound particle, which refers to the minimum binding energy 
particle. For two subhaloes $H_a$ and $H_b$ at subsequent redshift $z_a$ and $z_b$ 
$(z_a > z_b)$, $H_a$ is defined as the progenitor of $H_b$ , if at least 
half of the particles of $H_a$ are contained in $H_b$ , and the most bound particle 
of $H_a$ is also contained in $H_b$. Note that according to this definition, 
in the constructed merger trees each subhalo can have several progenitors, 
but has only one descendant. From snapshot to snapshot, the most massive progenitors
construct the main branch history of the merger trees, which can be used to calculate 
the formation times of the current subhaloes at $z = 0$.

In order to obtain more reliable formation time of the subhalo from the merger 
trees in ELUCID simulation, in the following analysis, we only 
focus on subhaloes at $z = 0$ with mass larger than 
$10^{10.5} \msun$ containing at least $100$ dark matter particles.

\subsection{Galaxies from SDSS DR7}

The galaxy sample used in this paper is from the SDSS \citep{York2000}, which is one of the most successful 
surveys in the history of astronomy. Based on the multi-band imaging and spectroscopic survey SDSS
DR7 \citep{Abazajian2009}, \citet{Blanton2005} constructed the New York University Value-Added Galaxy 
Catalog (NYU-VAGC) with an independent set of improved reductions. From the NYU-VAGC, we collect a total 
of $639, 359$ galaxies with redshifts in the range $0.01 \leq z \leq 0.2$, with redshift completeness
${\cal C}_z > 0.7$, and with extinction-corrected apparent magnitude brighter than $r \leq 17.72$.

The absolute magnitudes of galaxies are calculated and K-corrected and evolution corrected to 
$z = 0.1$ \citep{Yang2007}, using the method described by \citet{Blanton_2003} and \citet{Blanton_2007}.
The stellar masses and star formation rates of galaxies are from the public catalog of 
\citet{Chang2015} with reliable aperture corrections. Besides, we have checked our results using 
the stellar mass computed by the relation between stellar mass-to-light ratio and the color of 
\citet[see also \citet{Yang2007} for details]{Bell_2003}. We found that the final results are 
not sensitive to the different stellar mass used. Throughout this paper, the stellar masses 
from \citet{Chang2015} are used unless stated otherwise.

A total of $639, 359$ galaxies have a sky coverage of $7, 748$ square degree, consisting of 
two parts: a large continuous region in the Northern Galactic Cap (NGC) and a small region
in the Southern Galactic Cap. Since the initial condition of the constrained ELUCID simulation
is constructed from the distribution of the galaxies in the continuous NGC region,
in the following analysis, we only select galaxies located in the range 
$99^{\circ} < \alpha < 283^\circ$ and $-7^{\circ}<\delta < 75^{\circ}$,
resulting in a total
of 396, 069 galaxies in the continuous NGC region. Here $\alpha$ and $\delta$ are the right 
ascension and declination, respectively.

\subsection{Linking galaxies in observation to subhaloes in simulation}

\begin{figure*}
\includegraphics[width=0.49\textwidth]{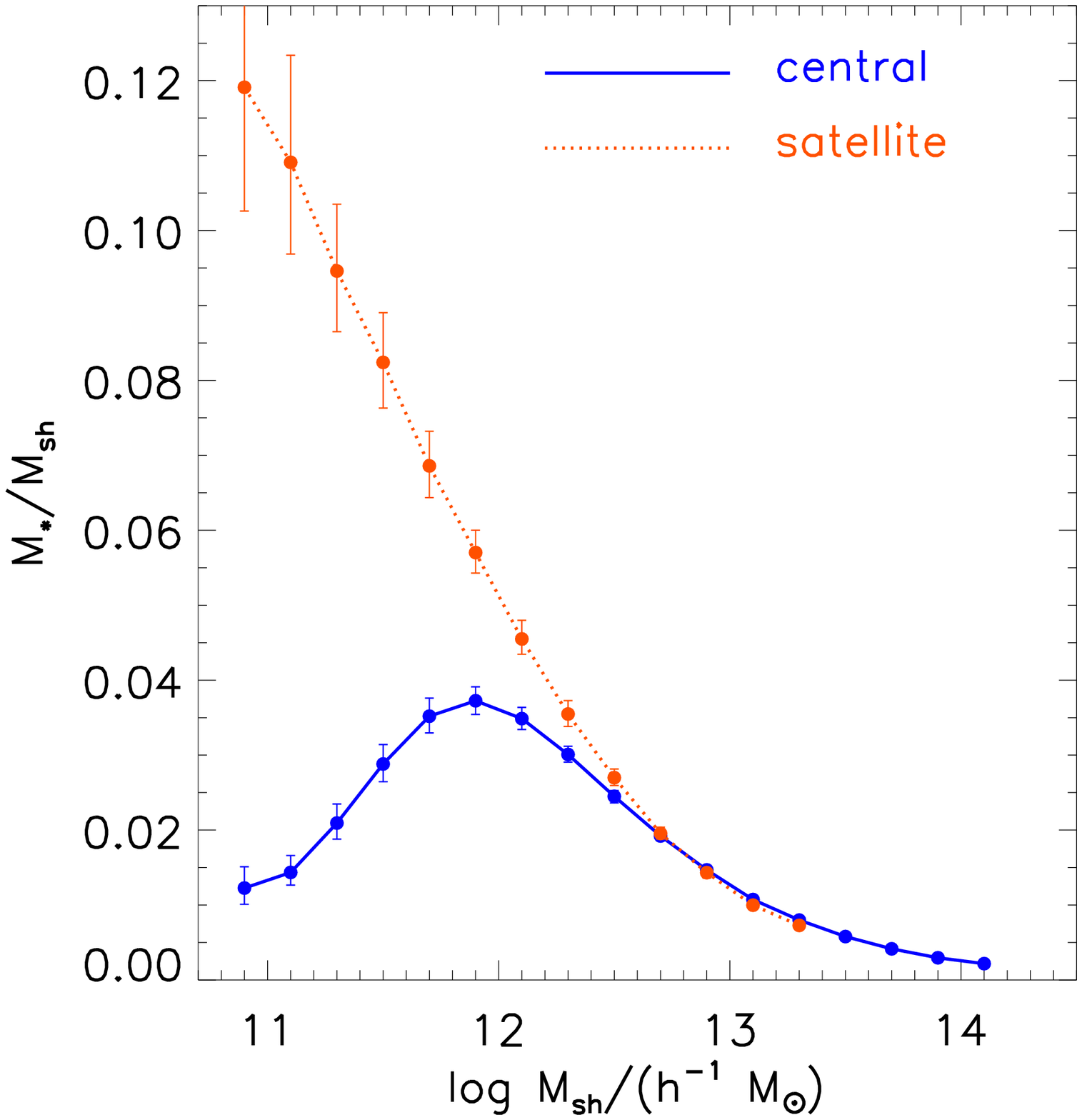}
\includegraphics[width=0.49\textwidth]{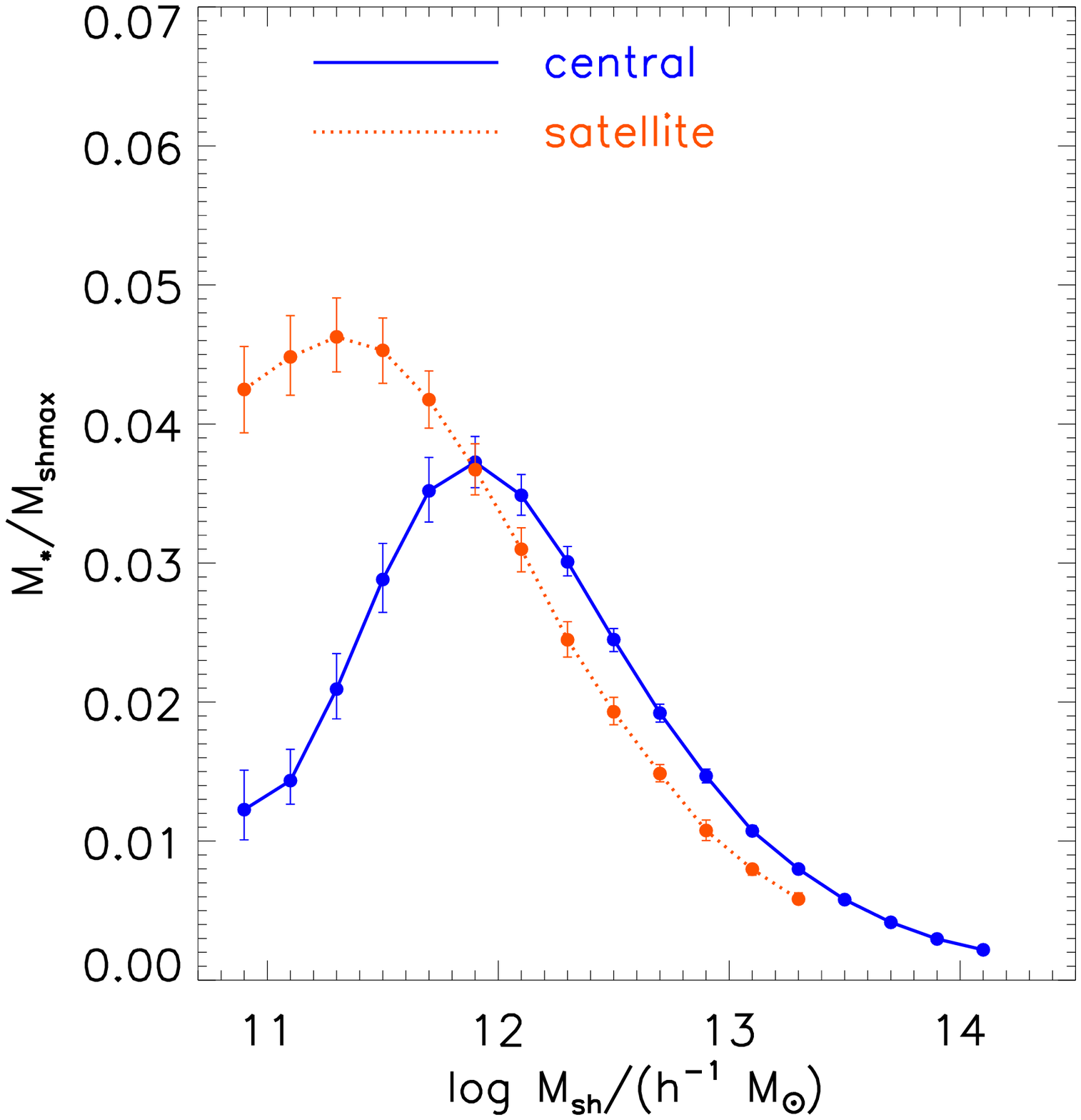}
\caption{The ratio of galaxy stellar mass relative to the current (left panel) and maximum (right panel) 
subhalo mass as a function of the subhalo mass. The blue solid and red dashed lines show the median values
for central and satellite galaxies, respectively. The error bars show the $45$ per cent and $55$ per cent
quantiles.}
\label{fig:shmr_mass}
\end{figure*}

In the constrained ELUCID simulation, the spatial distributions of the subhaloes at present are 
tightly correlated with the distributions of galaxies in SDSS DR7 \citep{WangHuiyuan2016,Yang2018},
because the simulation makes use of the initial condition constrained by the mass density field
extracted from the distribution of galaxies in SDSS DR7.

Galaxies in observation are linked to dark matter subhaloes in the ELUCID simulation using 
a neighborhood abundance matching method \citep{Yang2018}. For each galaxy in observation, 
we search for its corresponding subhalo in simulation according to the likelihood
\begin{equation}\label{eqn:match} 
P = M_{\rm shmax} \exp \left( - \frac {r_p^2} {2 r_{\rm off}^2} \right) 
\exp \left( -\frac {\pi^2} {2 v_{\rm off}^2}  \right), \end{equation}
where $r_p$ and $\pi$ are the separations in the redshift space between the galaxy and 
the subhalo in the perpendicular and parallel to the line-of-sight directions, 
respectively. $r_{\rm off}$ and $v_{\rm off}$ are two free parameters, while $M_{\rm shmax}$
is the peak mass of the subhalo under consideration. For $r_{\rm off}$ and $v_{\rm off}$ , 
the parameters are set to be $r_{\rm off} = 5 \mpc$ and $v_{\rm off} = 1000 ~{\rm km/s}$, 
which can give better constraint for the stellar-to-subhalo mass relation, especially
for the low-mass subhaloes. Note that in Equation~\ref{eqn:match}, the mass $M_{\rm shmax}$ 
is the dominant variable in the neighborhood abundance matching method, which will
degrade to the traditional abundance matching method if $r_{\rm off} = \infty$ and 
$v_{\rm off} = \infty$ in the extreme case.

Generally, the subhalo abundance matching models populate subhaloes with galaxies base 
on the peak values of the mass or the circular velocity over the merger histories of the 
subhaloes \citep{Conroy2006, Moster2010,Reddick2013,Matt2017, Campbell2018, vdBosch2018},
since the maximum values are found to be better correlated with the galaxy clustering statistics
than the current values at redshift $z = 0$. The current and peak values of the subhaloes are
different in the accretion history, especially for satellite subhaloes. In the accretion and
merger history, satellite subhaloes are commonly subject to tidal stripping in a larger system 
and remove dark matter mass, while the stellar components in the core are more tightly bound
than dark matter and change the stellar masses slightly. Therefore, the peak values are better
tracers of the potential well that shapes the galaxy statistical properties.

In the neighborhood abundance matching method, we use the peak mass of the subhaloes to link 
central subhaloes in simulation to central galaxies in observation, and satellite subhaloes
to satellite galaxies, respectively. From the galaxy-subhalo connection catalog, 
Figure~\ref{fig:shmr_mass} shows the ratios of galaxy stellar mass $M_*$ to the current $M_{\rm sh}$ 
and peak $M_{\rm shmax}$ subhalo mass for central and satellite galaxies, respectively. Here the galaxy
stellar mass is estimated from \citet{Chang2015},  and the subhalo mass is obtained 
from the matched subhalo in ELUCID simulation using the neighborhood abundance matching method. 

Figure~\ref{fig:shmr_mass} shows that central and satellite galaxies have different stellar-to-subhalo
mass ratios $M_*/M_{\rm sh}$ at fixed subhalo mass, especially for galaxies in low-mass subhaloes.
This is due to the different formation and evolution histories between central and satellite galaxies.
Satellite subhaloes commonly lose their mass due to tidal stripping after accretion into a larger system, 
while the concentrated stellar components change their mass slightly in the core of the subhalo. 
This behavior can be also indicated by comparing the red dashed lines between the left and right panels.
For satellites at fixed subhalo mass, the ratio $M_*/M_{\rm sh}$ is significantly larger than the ratio 
$M_*/M_{\rm shmax}$, especially for the low-mass satellite subhaloes, which indicates that satellite galaxies 
in low-mass subhaloes lose their dark matter mass significantly.

However, for centrals, the blue solid lines in left and right panels show that the ratios 
$M_*/M_{\rm sh}$ and $M_*/M_{\rm shmax}$ at fixed subhalo mass have almost no difference, since for 
centrals the maximum mass $M_{\rm shmax}$ is almost the same as the current mass $M_{\rm sh}$ at
$z=0$.

\begin{figure}
\includegraphics[width=0.5\textwidth]{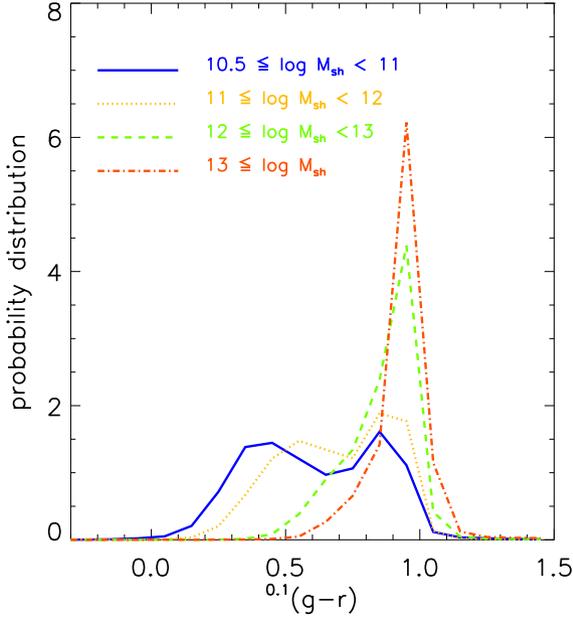}
\caption{Probability distribution function of the colors of the
galaxies within subhaloes in different mass ranges, which
are indicated by different type lines in the figure.}
\label{fig:color_pdf}
\end{figure}

Given each galaxy in observation is now linked with a subhalo in the ELUCID simulation, we show in 
Figure~\ref{fig:color_pdf} the probability distribution of the colors $^{0.1}(g-r)$ of 
galaxies within the fixed subhalo mass bins. In the galaxy-subhalo
connection catalog, there are $42, 431$, $236, 836$, $87, 541$ and $9, 813$ galaxies with
their corresponding subhalo mass $\log(M_{\rm sh}/(\msun))$ in the ranges of $(10.5, 11)$, $(11, 12)$, 
$(12, 13)$, and $(13, \infty)$, which are indicated by different type lines in Figure \ref{fig:color_pdf}.
Obviously, the galaxies in the subhaloes with $\log(M_{\rm sh}/(\msun))$ mass in the ranges $(10.5,11)$ and $(11,12)$
have the bimodal distribution in the $^{0.1}(g-r)$ color. This behavior is similar to the bimodal distribution
of galaxy colors at fixed absolute magnitude \citep{Baldry2004}. 

\subsection{subhalo formation time and environment}

\begin{figure}
\includegraphics[width=0.5\textwidth]{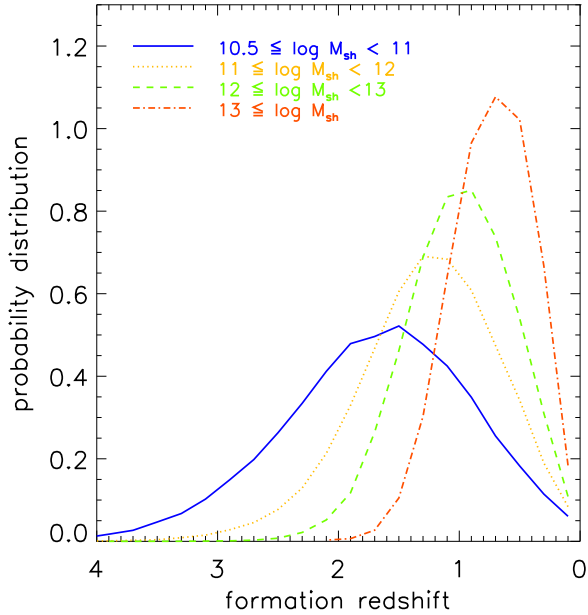}
\caption{Probability distribution function of the formation redshifts of 
subhaloes in different mass ranges.}
\label{fig:age_pdf}
\end{figure}

\begin{figure}
\includegraphics[width=0.5\textwidth]{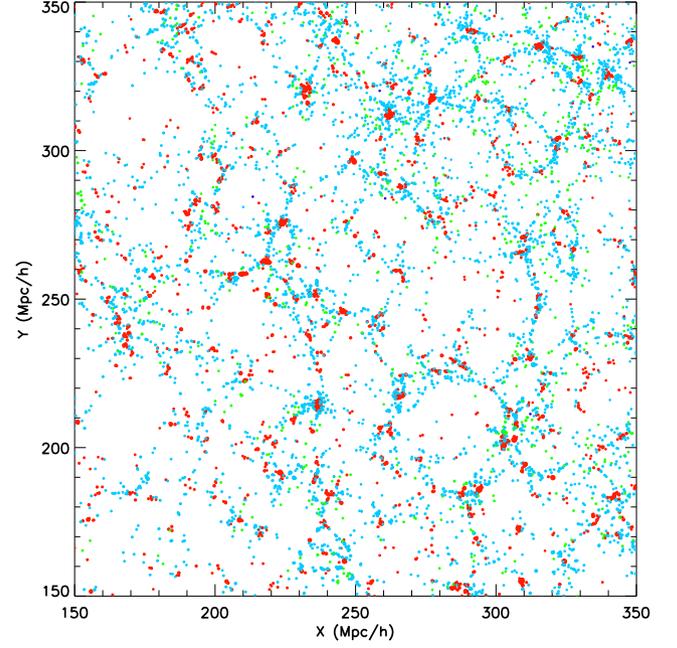}
\caption{Spatial distribution of subhaloes in different environments in a slice of 
thickness $10 \mpc$. Only dark matter subhaloes linked to galaxies in SDSS DR7 are shown. 
The symbol sizes of the subhaloes are proportional to their radii. The subhaloes in 
four different environments are indicated by different colors: \knot (red), \filament (cyan),
\sheet (green) and \void (blue).}
\label{fig:slice}
\end{figure}

Based on subhaloes identified by the algorithm SUBFIND, merger trees are constructed by linking a 
subhalo in one snapshot to a single descendant subhalo in the subsequent snapshot. Therefore, 
the halo merger tree is indeed a subhalo merger tree in this study \citep{Springel2005a}. The formation time of the subhalo
is defined as the time at which the main branch progenitor reached half of its maximum 
mass $M_{\rm shmax}$. 

Figure~\ref{fig:age_pdf} shows the probability distribution 
of the formation redshift of the subhaloes in different mass ranges, which are indicated 
by different type lines in the figure. Here again, we separate the total of $396,069$ subhaloes in the galaxy-subhalo pairs into four subsamples
according to their mass with $\log(M_{\rm sh}/(\msun))$ in the ranges $(10.5, 11)$, $(11, 12)$, 
$(12, 13)$, and $(13, \infty)$. For $236, 836$ subhaloes in the mass range
$10^{11} \msun \le M_{\rm sh} < 10^{12} \msun$, about $68\%$ subhaloes have their formation 
time in the redshift range $0.65 \leq z \leq 1.75$.

The geometric environment in which the subhalo (galaxy) resides is classified into \knot, 
\filament, \sheet, and \void~ by the Hessian matrix of the smoothing density field
\citep{Zhang2009}
\begin{equation}
H_{ij}(\bx) = \frac {\partial^2 \rho_s ({\boldsymbol x}) } 
{\partial x_i \partial x_j},
\end{equation}
where $i$ and $j$ are the Hessian matrix indices with values of $1$, $2$ or $3$. The smoothing 
density field $\rho_s$ is  calculated by the Gaussian filter with a fixed smoothing scale
$R_s = 2.1 \mpc$ \citep{Arag2007, Hahn2007b, Zhang2009}. The eigenvalues of the Hessian
matrix $H_{ij}$ are calculated at the position $\boldsymbol x$ of each subhalo. According to 
the number of negative eigenvalues, the subhalo’s environment is classified into one of 
four cosmic web types. If all of the three eigenvalues are negative, the subhalo’s environment is 
classified into \knot, while the case of two, one or zero negative eigenvalue corresponds to
\filament, \sheet ~or \void, respectively \citep{Zhang2009}.

Of the $396, 069$ subhaloes (galaxies) in galaxy-subhalo connection catalog, $148, 055$ $(37.4\%)$ 
are located in \knot, $217, 746$ $(55.0\%)$ are located in \filament, $29, 909$ $(7.6\%)$
are located in \sheet ~and $359$ $(0.09\%)$ are located in \void. Figure~\ref{fig:slice} shows the 
spatial distribution of subhaloes in a $200 \mpc  \times 200 \mpc$ slice of thickness $10 \mpc$. 
Note that only subhaloes linked to galaxies in SDSS DR7 are shown in Figure~\ref{fig:slice}. 
The subhaloes in four different environments are indicated by different colors: \knot (red),
\filament (cyan), \sheet (green) and \void (blue).

\section{Proxy of halo formation time}\label{sec_proxy}

In this section, we investigate how the galaxy properties are correlated with subhalo formation time,
using the catalog of galaxies linked to the subhaloes in the ELUCID simulation. To find a reasonable
proxy of subhalo formation time, we mainly focus on the ratio of stellar mass 
to subhalo mass. In most of our subsequent probes, we separate central or satellite galaxies into four subsamples with their corresponding subhalo 
mass $\log ( M_{\rm sh}/(\msun))$ in the ranges of $(10.5, 11)$, $(11, 12)$,  $(12, 13)$, and $(13, \infty)$.

\subsection{formation redshifts for central and satellite subhaloes}
\begin{figure}
\includegraphics[width=0.5\textwidth]{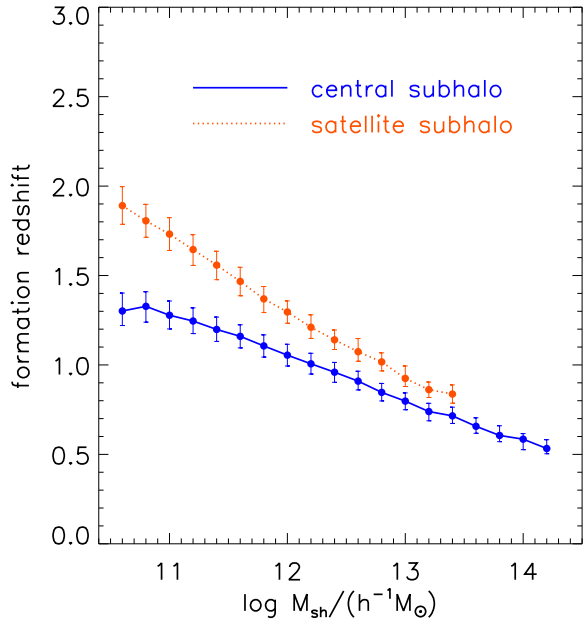}
\caption{Subhalo formation redshifts as a function of the mass of central and satellite 
subhaloes. The blue solid and red dashed lines show the median values for central and
satellite subhaloes, respectively.}
\label{fig:age_mass}
\end{figure}

In simulation, we define central subhalo to be the most massive subhalo in a given host FOF halo, and
satellite subhaloes to be any other subhaloes. Based on the subhalo merger trees from ELUCID simulation,
the formation time of the subhalo is defined as the redshift at which the main branch progenitor 
reached half of its maximum mass $M_{\rm shmax}$. To obtain a reliable measurement of the subhalo 
formation time, we mainly focus on the subhaloes containing at least $100$ dark matter particles
with mass larger than $10^{10.5}\msun$.

In the evolutionary process from high redshift to the present, the mass of central subhalo generally 
grows larger and larger across the main branch of the merger trees, while satellite subhaloes usually
lose their mass on account of tidal stripping or dynamical friction in the process of accretion into 
the host haloes \citep{vdBosch2018,Rod2020}. Therefore, the distinct assembly histories of central and satellite subhaloes result in different
subhalo formation time for central and satellite subhaloes. Figure~\ref{fig:age_mass} shows the median
values of the formation redshifts of central and satellite subhaloes as a function the subhalo mass.
Obviously, at fixed subhalo mass, the formation redshifts of satellite subhaloes are larger than
those of central subhaloes. 

Therefore, in the following analysis, we investigate the properties of central and satellite subhaloes 
separately, due to the distinct assembly history of central and satellite subhaloes.

\subsection{galaxy stellar-to-subhalo mass ratio}

\begin{figure}
\includegraphics[width=0.49\textwidth]{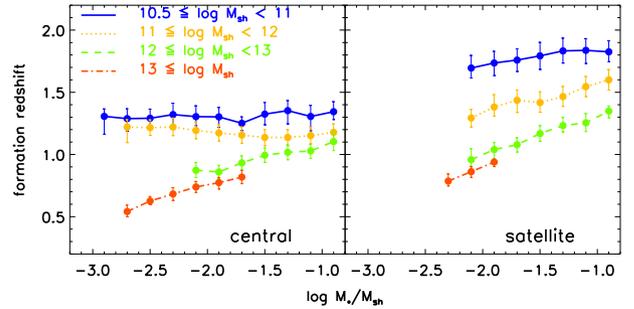}
\caption{Subhalo formation redshift as a function of the stellar-to-subhalo mass ratio for central and satellite galaxies in different subhalo mass ranges.}
\label{fig:z_ratio}
\end{figure}

The galaxy stellar-to-halo mass relation has been extensively studied using various techniques, such as
the measurements of galaxy-galaxy lensing \citep{Mandelbaum2016}, satellite kinematics \citep{Lange2019}, and 
abundance matching model \citep{Chaves2016, Dragomir2018, Contreras2020}. 

With the help of the constrained ELUCID simulation, we can directly make a one-to-one comparison between galaxy stellar-to-subhalo
mass ratio and the formation time of the subhalo in simulation. Using the neighborhood subhalo abundance matching
method to the constrained simulation, galaxies in observation are linked to subhaloes in ELUCID simulation, according to not only their mass, 
but also their positions. Therefore, the galaxy-subhalo connection catalog used in this paper is more reliable than that
generated by traditional abundance matching model.

Figure~\ref{fig:z_ratio} shows the formation time of the subhalo as a function of galaxy stellar-to-subhalo mass ratio $M_*/M_{\rm sh}$ for central and satellite galaxies. As 
shown in the left panel of Figure~\ref{fig:z_ratio}, 
the subhalo formation redshifts increase with the stellar-to-subhalo mass ratios for central galaxies in 
subhaloes with $M_{\rm sh} \ge 10^{12} \msun$, while for central galaxies with subhalo
mass $M_{\rm sh} < 10^{12} \msun$, the formation time of the subhalo seems to be independent on galaxy stellar-to-subhalo 
mass ratio. The flat trend blue and orange lines in the left panel of Figure 6 may be because 
the matched galaxy-subhalo pairs are dominated by Poisson errors \citep{Tweed2017} and less reliable for low-mass subhaloes. 
As we have checked the mass ratios of the second to most massive subhaloes around the central galaxies, the ratio for
subhaloes with mass $<  10^{12}h^{-1}M_\odot$ is at $\ga 0.8$ level. Because of the
scatter in the stellar-subhalo mass relation, it is not necessary that the most massive subhaloes give  the best 
match with the central galaxies for these low mass subhaloes. While for those subhaloes with mass $\ga 10^{12}h^{-1}M_\odot$, 
the match should be much more reliable. 
In addition to this, in galaxy-subhalo pairs, there are a fraction of matched pairs with very large 
separations \citep[see Figure. 2 in][]{Yang2018}, and the median projected and line-of-sight separations 
are $r_p=2.6 \mpc$ and $\pi=3.3 \mpc$, respectively. To check the impact of this, we have also
generated several closely-matched subsamples
with smaller separation $r_p$ and $\pi$, and performed similar measurements of Figure~\ref{fig:z_ratio}.
We find that the trends of closely-matched subsamples keep almost unchanged compared with the current 
results.

In addition, we investigate the formation time of satellite galaxies as a function of galaxy stellar-to-subhalo mass
ratio. For satellites galaxies as shown in the right panel of Figure~\ref{fig:z_ratio}, 
the subhalo formation redshifts also increase with the stellar-to-subhalo mass ratio.

Note that the initial condition of the constrained simulation is constructed using the galaxy groups with
mass larger than $10^{12} \msun$, thus the galaxy-subhalo pairs with subhalo mass larger than  $10^{12} \msun$
are more reliable in this study. In a word, the galaxy stellar-to-subhalo mass ratio is a good proxy of 
the formation time, especially for galaxies in high-mass subhaloes.

\begin{figure}
\includegraphics[width=0.49\textwidth]{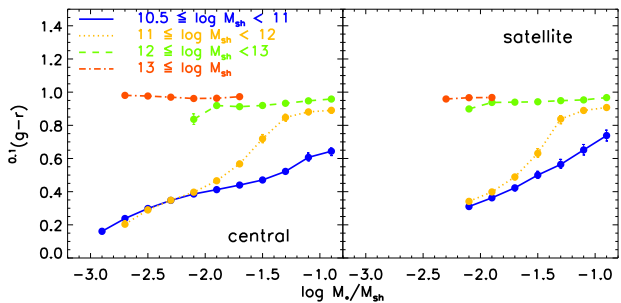}
\caption{Galaxy colors as a function of the stellar-to-subhalo mass ratio for central
and satellite galaxies. The solid points with different types of lines show the median values for galaxies
in different subhalo mass ranges.  
}
\label{fig:color_ratio}
\end{figure}

\subsection{galaxy color}

Using N-body simulations,  \citet{WangH2011} investigated the correlations among different halo properties, and found that the mass ratio 
between the most massive subhalo and its host FOF halo is tightly correlated with the halo formation time, spin and shape. 
Inspired by this, \citet{Lim2016} investigated the correlation between galaxy stellar-to-halo mass ratio and galaxy properties, 
using galaxy groups selected by the halo-based group finder \citep{Yang2007} from SDSS DR7. They found that galaxy 
stellar-to-halo mass ratio is tightly correlated with galaxy colors and star formation in observation.

Using the 396,069 galaxy-subhalo pairs, here we check the relation between the galaxy stellar-to-subhalo mass ratio and galaxy 
colors. Figure~\ref{fig:color_ratio} shows the median values of galaxy colors as a function of the ratio of galaxy 
stellar-to-subhalo mass for central and satellite galaxies in different subhalo mass ranges,
indicated by different types of lines in the figure. For galaxies in subhaloes with mass less than $10^{12} \msun$, the galaxy stellar-to-subhalo mass ratio 
is tightly correlated with the galaxy colors, where redder galaxies have higher stellar-to-subhalo mass ratios. 
Note that, for a given subhalo mass, the galaxy stellar-to-subhalo mass ratio is proportional to the galaxy stellar mass. 
In general, the color dependence seen here might be attributed mainly by the stellar mass-color dependence. 
Only using observational data, \citet{Lim2016} also found that redder galaxies have higher stellar-to-group mass
ratios, where the groups are selected by the halo-based group finder \citep{Yang2007}. In this study,
using closely-matched subsamples with smaller separation $r_p$ and $\pi$, we have confirmed the color dependence of 
galaxy stellar-to-subhalo mass ratio for galaxies in low-mass subhaloes. The blue and orange lines in Figure~\ref{fig:color_ratio}
show that lower stellar-to-subhalo mass ratio galaxies are bluer, which seems to indicate that their star formation
started later. Although because of the potential mismatches between low mass subhaloes and central galaxies, where 
the correlation between the subhalo formation time and the stellar-to-subhalo mass ratio can not be well 
recovered, the satellite galaxies do indicate that the galaxy colors are correlated with the subhalo formation time. 

For galaxies in more massive subhaloes with mass larger than $10^{12} \msun$, there is 
almost no dependence of galaxy colors on the stellar-to-subhalo mass ratio, because galaxies are all equally red in the high mass range.

\section{Dependence on the cosmic web}\label{sec_cw}

The cosmic web environment in which the subhalo (galaxy) resides is classified into konts, 
filaments, sheets, and voids by the Hessian matrix of the smoothing density field \citep{Zhang2009}.
Based on the geometric cosmic web classification, we investigate the dependence of subhalo formation time
and galaxy properties on the cosmic web environment, using the catalog of galaxies in observation 
and subhaloes in ELUCID simulation. 

\begin{figure}
\includegraphics[width=0.49\textwidth]{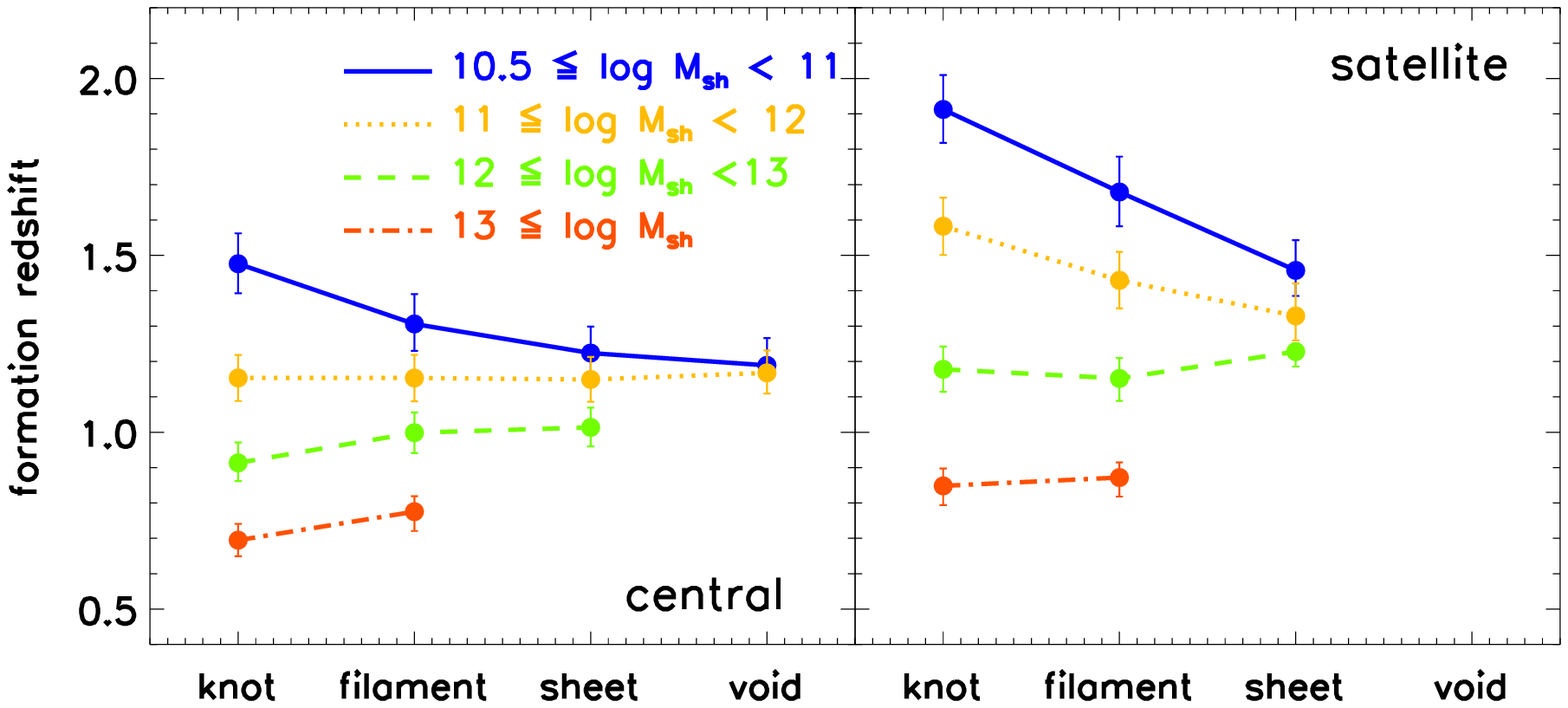}
\includegraphics[width=0.49\textwidth]{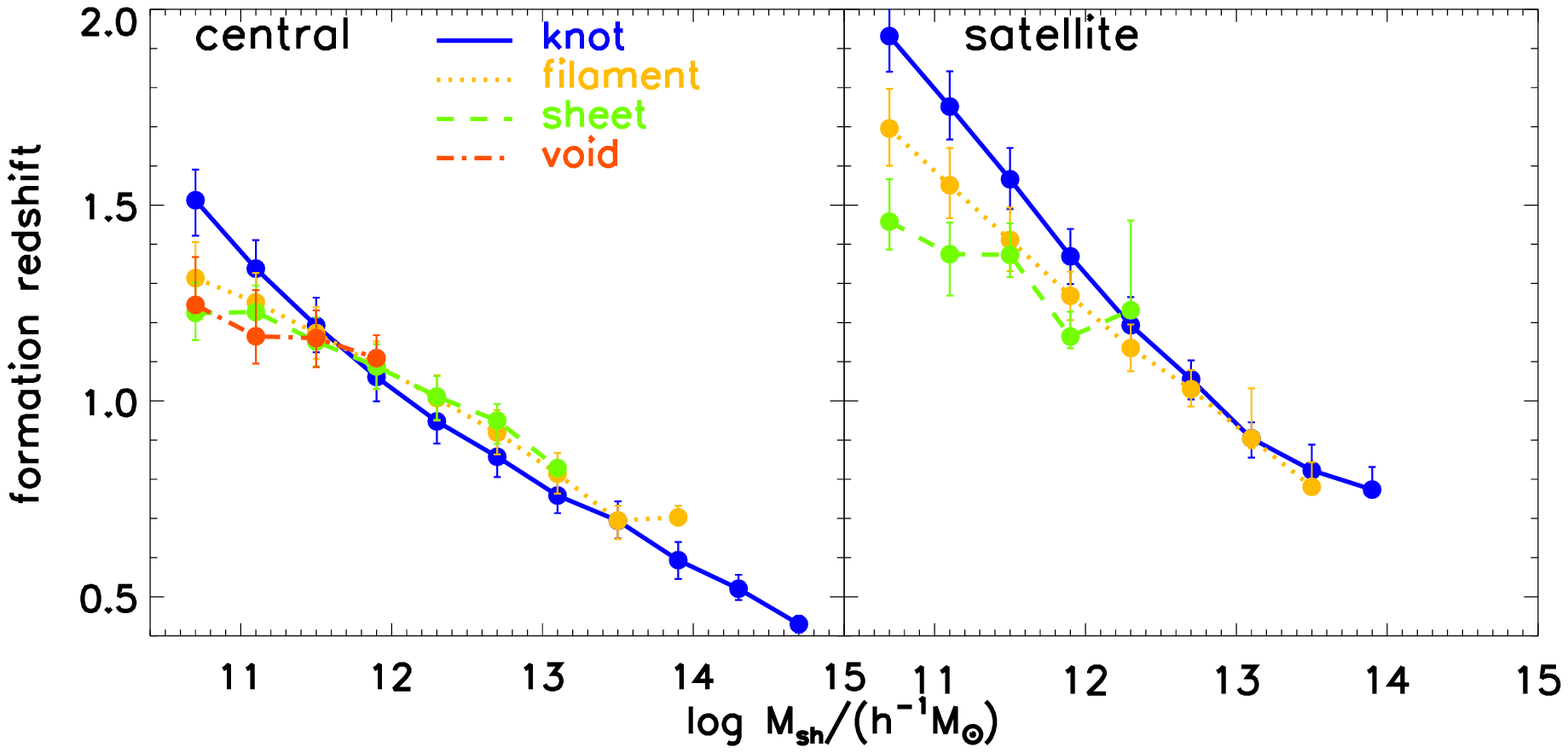}
\caption{Subhalo formation redshift as a function of the cosmic web environment (top panels) and the subhalo mass (low panels) for central
and satellite subhaloes. For satellite subhaloes in the right panels, there are no data points in the \void, because there are 
almost no satellite subhaloes in the \void~ in our sample.}
\label{fig:z_cw}
\end{figure}

\subsection{subhalo formation time}

Using subhaloes from the ELUCID simulation, we investigate the dependence of the subhalo formation time on 
the cosmic web environment. The formation redshift of the subhalo is defined as the redshift
at which the main branch progenitor reached half of its maximum mass $M_{\rm shmax}$.

In the top panels of Figure~\ref{fig:z_cw}, we show the subhalo formation redshifts as a function of
the cosmic web environment for central and satellite subhaloes in different mass range. For central
subhaloes, in the low-mass range $(10.5, 11)$, subhaloes in knots are older than subhaloes of the same mass in
filaments, sheets or voids, while in the high-mass range $(12, 13)$ and $(13, \infty)$, subhaloes in knots
are younger than haloes of the same mass in sheets or filaments. For clarity, the low panels of Figure~\ref{fig:z_cw} 
shows the subhalo formation redshifts as a function of the subhalo mass in different cosmic environments.
For central subhaloes, there is a characteristic subhalo mass of $10^{11.7} \msun$, below which the formation
redshifts are larger in knots than filaments, sheets or voids, while above which it reverses. For central 
subhaloes at high mass, the formation redshifts are smaller in knots. 

As shown in the right panels in Figure~\ref{fig:z_cw}, for satellite subhaloes below $10^{12.5} \msun$, subhaloes
in knots are older than subhaloes of the same mass in filaments or sheets. Above $10^{12.5} \msun$,  there are only 
$2$ satellite subhaloes in sheets and no satellite subhaloes in voids. Thus, the results are not robust for satellite
subhaloes in high-mass range.

Actually, here the environmental dependence 
of the subhalo formation time is independent of the matching method, because the subhalo formation time, subhalo mass 
and cosmic web classifications used in Figure~\ref{fig:z_cw} are all from the simulation data. We have confirmed that
similar results are obtained for a total of subhaloes including unmatched pairs in the ELUCID simulation. 
The environmental dependence of the halo formation time below or above the characteristic subhalo mass of $\sim 10^{12} \msun$ 
that we have found is consistent with the analysis of halo assembly bias. For low-mass haloes, older haloes cluster more
strongly than younger haloes \citep{Gao2005, Gao2007}, while for high-mass haloes, the trend is reversed, where older haloes are 
less clustered \citep{Jing2007}. One explanation is that for low-mass haloes in the knots environment, they are relatively located in
the vicinity of clusters \citep{Hahn2007a, Hahn2007b}, and their mass accretions are suppressed by the hot environments produced by the 
tidal fields of clusters \citep{WangH2007}, resulting in the older ages of low-mass haloes in knots. However, for high-mass haloes, 
the heating of the large-scale tidal field is less important, and haloes in knots grows more easily by continuous accretion and major mergers.

To check if the above found features are robust with respect to our construction of the cosmic web, we perform our 
analysis for cosmic web constructed with additional two choices of smoothing scales  $R_s=4.2$ and $8.4\mpc$. 
We find that the transition characteristic subhalo mass for central subhalos slightly increases with the increase of the smoothing 
scale we used, as there are relatively more massive subhaloes classified into void regions. Nevertheless, 
the overall trends are very similar.

\begin{figure}
\includegraphics[width=0.49\textwidth]{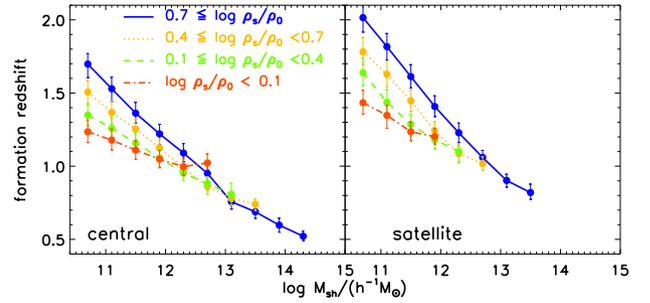}
\caption{Subhalo fromation redshift as a function of the subhalo mass for central
and satellite subhaloes in different cosmic density ranges of $\log \rho_s/\rho_0$, where
$\rho_s$ is the smoothing density with Gaussian filter of $R_s = 2.1 \mpc$ at the 
location of a given subhalo, and $\rho_0$ is the mean cosmic density of the universe.}
\label{fig:age_density}
\end{figure}

\begin{figure}
\includegraphics[width=0.49\textwidth]{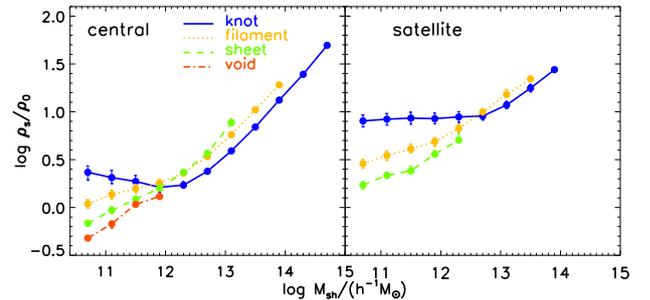}
\caption{Cosmic density field as a function of the subhalo mass in different cosmic web environments for central
and satellite subhaloes. }
\label{fig:density_mass}
\end{figure}

\begin{figure}
\includegraphics[width=0.49\textwidth]{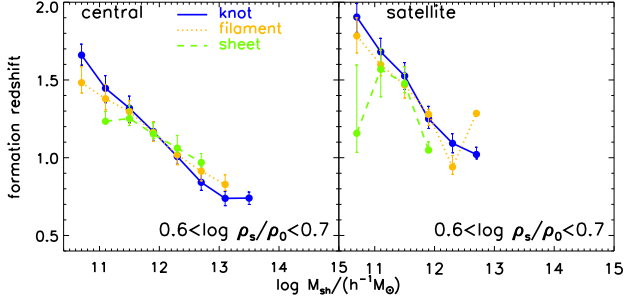}
\caption{Similar to the low panels of Figure~\ref{fig:z_cw}, but for subhaloes in the cosmic density range of
$0.6<\log \rho_s/\rho_0 <0.7$.}
\label{fig:age06}
\end{figure}

In order to further understand the environmental dependence of the subhalo formation time, we also investigate the 
effects of the local environmental density on the subhalo formation time. At the location of a given subhalo, the smoothing
density field $\rho_s$ is calculated by smoothing the cloud-in-cell (CIC) generated density field with a spherically 
symmetric Gaussian filter of $R_s = 2.1 \mpc$ \citep{Zhang2009}. Then we separate central or satellite subhaloes into four
equal subsamples with their local environmental density $\log \rho_s/\rho_0$ in the ranges of $(-0.8, 0.1)$, $(0.1, 0.4)$,  
$(0.4, 0.7)$, and $(0.7, 2.0)$, where $\rho_0$ is the average cosmic matter density of the universe. Figure~\ref{fig:age_density}
shows the subhalo fromation redshift as a function of the subhalo mass for central and satellite subhaloes in different 
environmental density. For centrals or satellites, subhaloes in higher density are older than those in lower density, especially 
for low-mass subhaloes. Figure~\ref{fig:density_mass} shows the cosmic density as a function of the subhalo mass in different
cosmic web environments. Obviously, for central subhaloes in the left panel of Figure~\ref{fig:density_mass}, 
there is a characteristic subhalo mass of $\sim 10^{12} \msun$, below which the cosmic density is larger in knots than in filaments,
sheets or voids, while above which it reverses. Combining the results of Figure~\ref{fig:age_density} and Figure~\ref{fig:density_mass},
it is reasonable that low-mass subhaloes in knots are denser and older than in other cosmic web environments and this trend reverses 
for high-mass subhaloes, which has been shown in Figure~\ref{fig:z_cw}.

In order to clarify the effect of the cosmic density and the cosmic web environment on the subhalo formation time, we 
generate $25$ subsamples keeping the similar environmental density $\log \rho_s/\rho_0$ from $-0.5$ to $2$, although in general subhaloes 
in different cosmic web environments have different cosmic density. We have investigated the cosmic web dependence of $25$ 
subsamples in the similar density with bin $\Delta \log \rho_s/\rho_0 = 0.1$. Figure~\ref{fig:age06} shows an example of the cosmic 
web dependence of the subhalo formation time in the cosmic density of $\log \rho_s/\rho_0$ in the range $(0.6,0.7)$. We find that
there still exists some cosmic web dependence of the subhalo formation time in the similar cosmic density for central or satellite subhaloes. 

\begin{figure}
\includegraphics[width=0.24\textwidth]{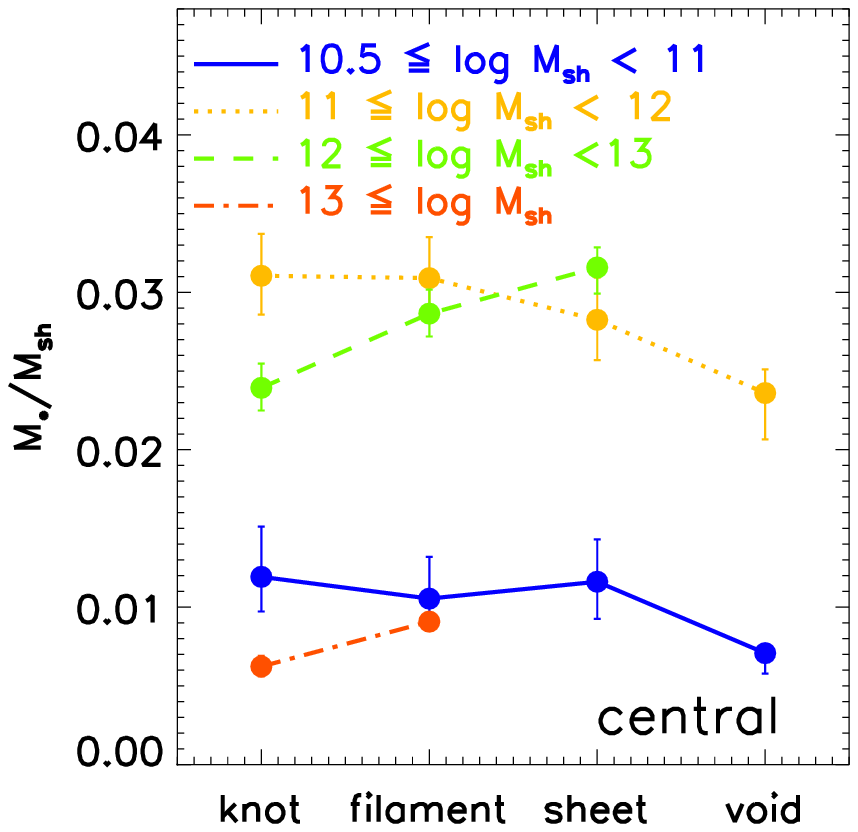}
\includegraphics[width=0.24\textwidth]{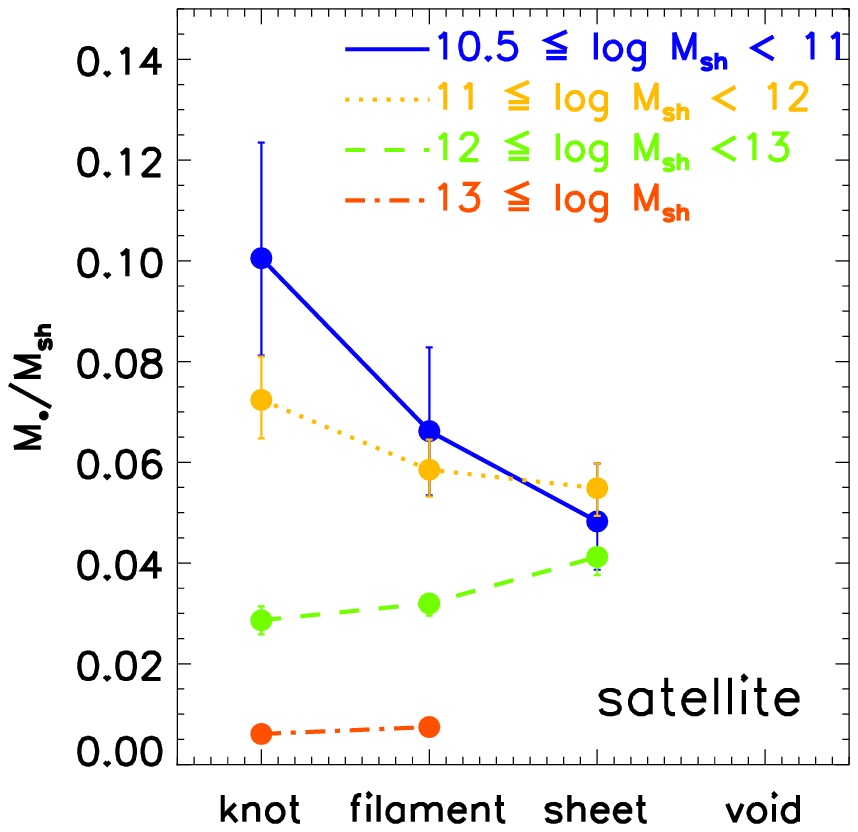}
\includegraphics[width=0.24\textwidth]{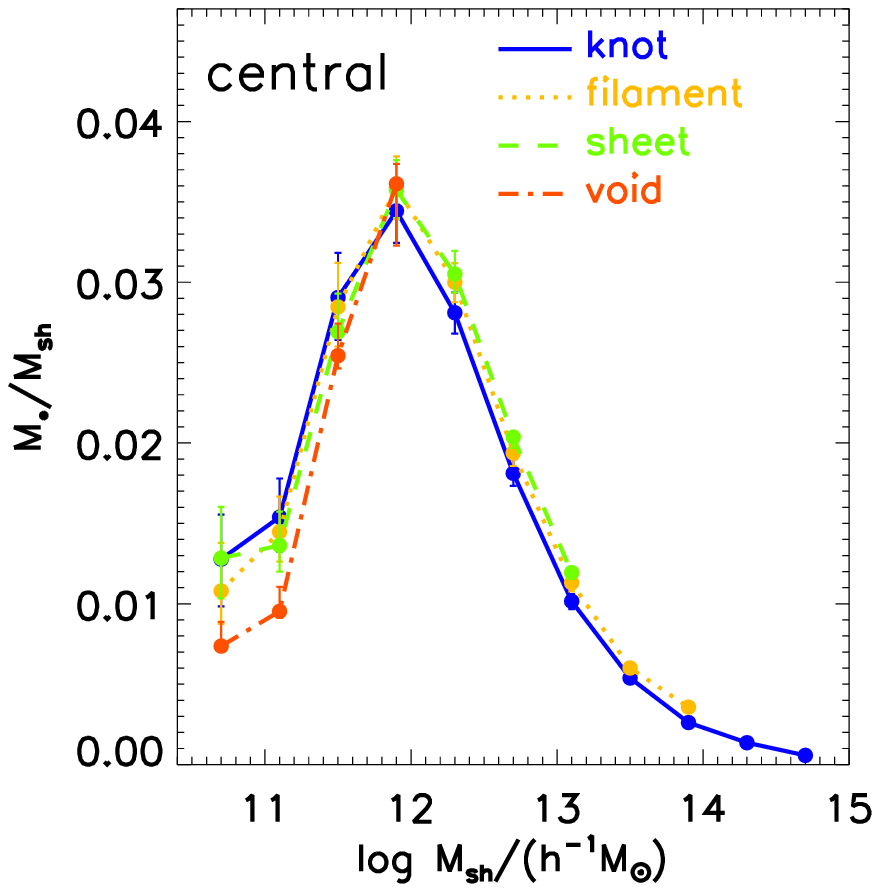}
\includegraphics[width=0.24\textwidth]{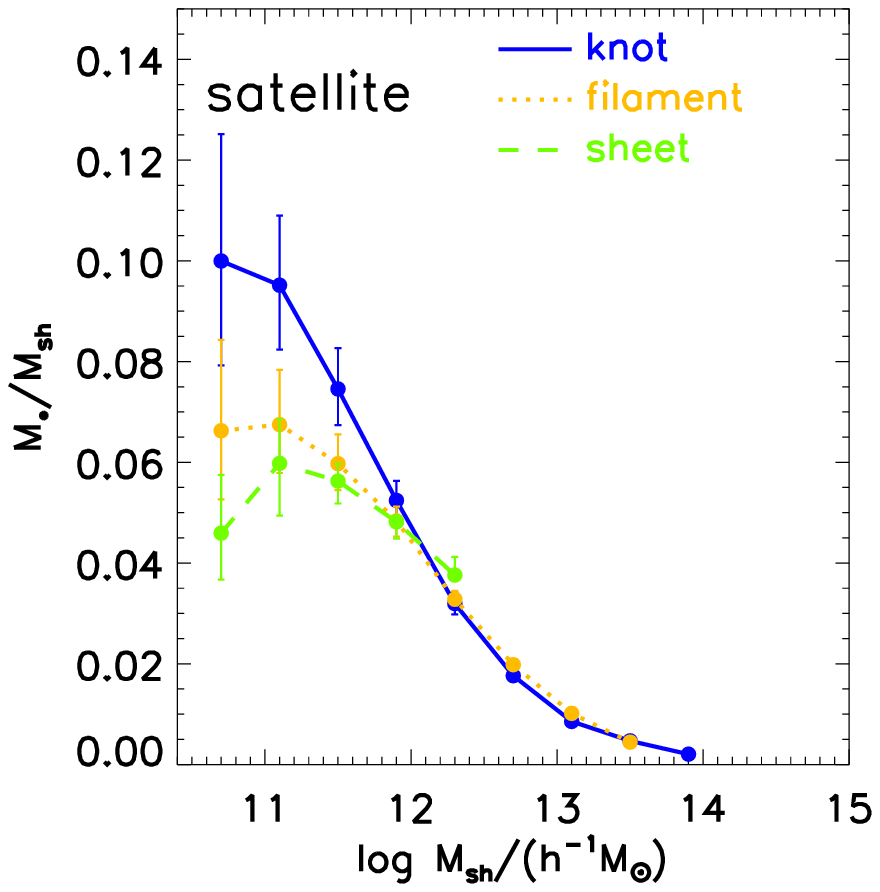}
\caption{Galaxy stellar-to-subhalo mass ratio as a function of the cosmic web environment (top panels) and the subhalo mass (low panels) for central
and satellite subhaloes. }
\label{fig:shmr_cw}
\end{figure}

\subsection{galaxy stellar-to-subhalo mass ratio }

The stellar masses of galaxies in SDSS are from the public catalog of \citet{Chang2015} . The subhalo masses are from the subhaloes
by the SUBFIND algorithm \citep{Springel2001} in ELUCID simulation. Linking galaxies in observation to subhaloes in simulation, we find that   
the galaxy stellar-to-subhalo mass ratio is a good proxy of the subhalo formation time, especially for galaxies in high-mass subhaloes.
Therefore, it's interesting to investigate the dependence of the galaxy stellar-to-subhalo mass ratio on the cosmic web environment.

The top panels of Figure~\ref{fig:shmr_cw} show the galaxy
stellar-to-subhalo mass ratio as a function of the cosmic web environment for central
and satellite galaxies, respectively. For central galaxies, in the mass ranges of 
$(10.5, 11)$ and $(11, 12)$, the stellar-to-subhalo mass ratios in knots are larger than those in filaments, sheets, or voids, while in the mass ranges of $(12, 13)$ and 
$(13, \infty)$, the ratios in knots are smaller than those in filaments or sheets.
For clarity, we also show the galaxy stellar-to-subhalo mass ratio as a function of 
the subhalo mass in different cosmic environment in the low panels of Figure~\ref{fig:shmr_cw}. For central galaxies, the stellar-to-subhalo mass ratio has a peak at the subhalo mass $\sim 10^{12}\msun$ and drops fast towards the low-mass or high-mass end. Similar to the dependence of the formation time on the cosmic web environment, there is a characteristic subhalo mass of $\sim 10^{12} \msun$, below
which the galaxy stellar-to-subhalo mass ratio is largest in konts, while above which
it reverses. 

For satellite galaxies, the dependence of the stellar-to-subhalo mass ratio on environment
is similar to that of central galaxies. For satellite galaxies with their corresponding
subhalo mass less than $\sim 10^{12}\msun$, the stellar-to-subhalo mass ratio decreases
from knots to filaments or sheets, while the ratio slightly increases from knots to
filaments with mass larger than $\sim 10^{12}\msun$. Note that no statistical results can
be obtained in voids for satellite galaxies due to the limited number.

The general trends of  the stellar-to-subhalo mass ratios on the cosmic web environment 
are quite consistent with those of the subhalo formation time, indicating that the former is 
indeed a good
tracer of the latter.

\begin{figure}
\includegraphics[width=0.49\textwidth]{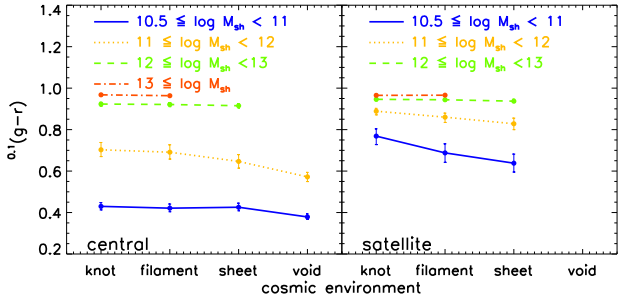}
\caption{Galaxy colors as a function of the cosmic web environment for central 
(left panel) and satellite (right panel) galaxies with different corresponding
subhalo mass $\log ( M_{\rm sh}/(\msun))$ in the ranges of $(10.5, 11)$, $(11, 12)$,
$(12, 13)$, and $(13, \infty)$.}
\label{fig:color_cw}
\end{figure}

\subsection{galaxy color and specific star formation rate}

Galaxy color is one of the most important variables in observation, and it's tightly
correlated with the galaxy specific star formation rate.  In this section, we mainly focus on the dependence of galaxy color and specific star formation rate on the cosmic web environment.

Figure~\ref{fig:color_cw} shows the median values of galaxy colors as a function of the cosmic web environment for central and satellite galaxies, respectively.  As expected, 
more massive galaxies have higher $^{0.1}(g-r)$ and tend to be redder.  For centrals or satellites with subhalo mass less than $10^{12} \msun$, galaxies in knots are redder than those in filaments, sheets, or voids. 
For galaxies in massive subhaloes with mass larger than $10^{12} \msun$, there is no clear difference of galaxy colors in different cosmic web environments, since the color difference is negligible for galaxies in high-mass subhaloes, and galaxies 
in massive subhaloes are all equally red as shown in Figure~\ref{fig:color_pdf}.

\begin{figure}
\includegraphics[width=0.49\textwidth]{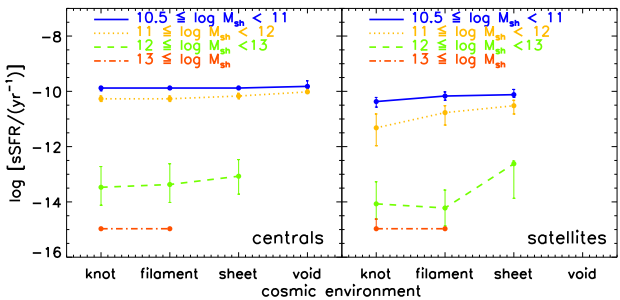}
\caption{Specific star formation rate as a function of the cosmic web environment
for central (left panel) and satellite (right panel) galaxies with different corresponding
subhalo mass $\log ( M_{\rm sh}/(\msun))$ in the ranges of $(10.5, 11)$, $(11, 12)$,
$(12, 13)$, and $(13, \infty)$.}
\label{fig:ssfr_cw}
\end{figure}

The specific star formation rates of galaxies are from the public catalog of \citet{Chang2015}. The specific star formation rate has been extensively used as the indicator of galaxy quiescence to separate the quenched from the star-forming population \citep{Favole2021}. 

Figure~\ref{fig:ssfr_cw} shows the median values of galaxy specific star formation rates as a function of the cosmic web environment for central and satellite galaxies, respectively.
For centrals or satellites, galaxies in knots have lower sSFRs than those in filaments, sheets, or voids, especially for galaxies with their corresponding subhalo mass less than
$10^{13} \msun$. This behavior is consistent with the environmental dependence of galaxy
colors.

\begin{figure}
\includegraphics[width=0.24\textwidth]{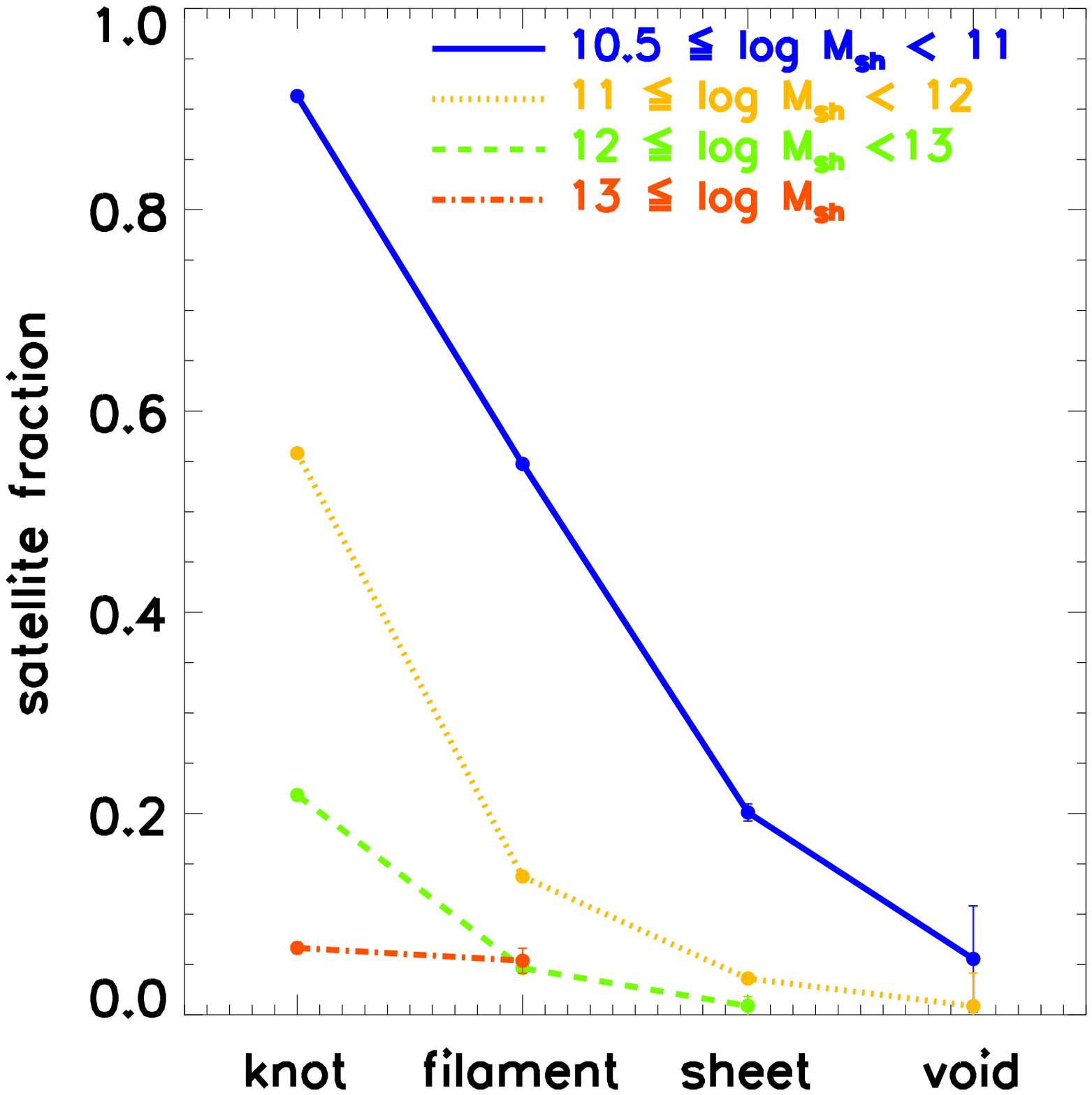}
\includegraphics[width=0.24\textwidth]{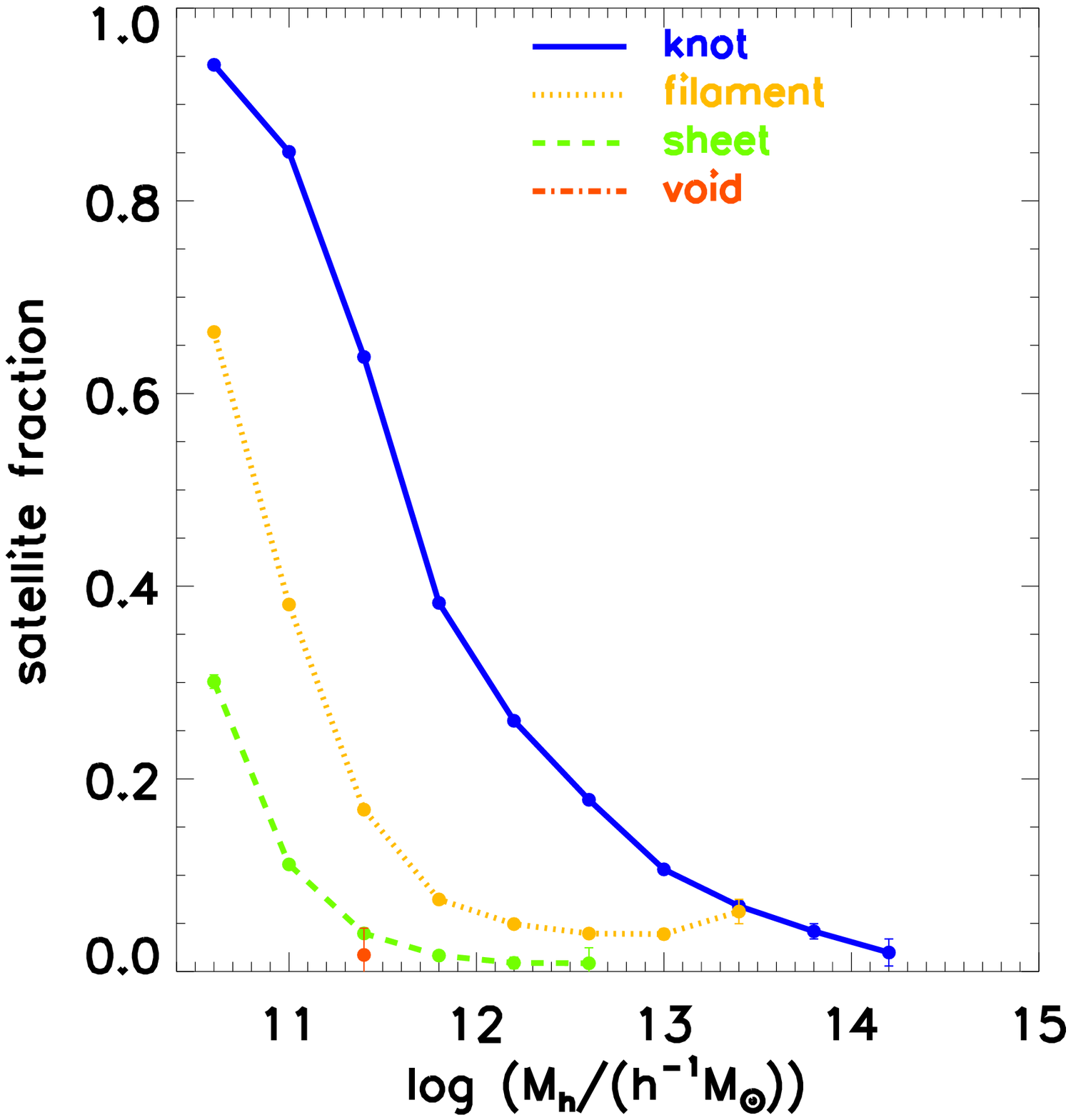}
\caption{ Satellite fraction as a function of the cosmic web environment (left panel) and the subhalo mass (right panel). }
\label{fig:frac_cw}
\end{figure}

\subsection{satellite fraction}

The satellite fraction of galaxies as a function of luminosity or stellar mass is important for understanding the clustering of galaxies at fixed luminosity or stellar mass 
\citep{Yang2012, Yang2018}, the measurements of galaxy-galaxy lensing signals \citep{Mandelbaum2006}, and the quenching 
of central and satellite galaxies \citep{Bluck2016}. Using galaxy-subhalo pairs catalog,
\citet{Yang2018} have shown that the satellite fraction decreasing with stellar mass, 
and red galaxies have significantly higher satellite fraction than bule galaxies, 
especially for low-mass galaxies.  Using galaxy-galaxy weak lensing from SDSS, 
\citet{Mandelbaum2006} also found early-type galaxies have higher satellite 
fraction than late-type galaxies at a give stellar mass.

Since galaxies in older subhaloes are more likely to be satellite galaxies, at a given mass, satellite fraction is tightly correlated with the subhalo formation time. 
In this section, we investigate the dependence of satellite fraction on the cosmic web environment.

The left panel of Figure~\ref{fig:frac_cw} shows the satellite fraction
as a function of the cosmic web environment for galaxies in different mass subhaloes. The satellite fraction decreases from knots
to filaments to sheets to voids, especially for low-mass galaxies. 
The right panel of Figure~\ref{fig:frac_cw} shows the satellite fraction as a 
function of the subhalo mass. Generally, the satellite fraction significantly 
decreases with increasing subhalo mass in different cosmic web environment. 
At fixed mass, galaxies in knots have larger satellite fraction than those in
filaments, sheets, or voids.

Since satellite fraction is strongly dependent on the cosmic web environment,
more satellites in knots can result in the redder colors in knots for a sample
of galaxies including centrals and satellites, considering that satellite galaxies
are redder than central galaxies at fixed mass. As a consequence, the properties of 
central and satellite galaxies are separately investigated in the previous sections. 
The cosmic web environmental dependence of galaxy colors will be significantly 
amplified for {\tt total} galaxies, if galaxies are not distinguished into central or satellite subsamples.

\section{Summary}\label{sec_summary}
In this paper, combining galaxies from SDSS DR7 and subhaloes from the ELUCID
simulation, we have investigated the relation between the formation time of subhaloes
and the galaxy properties. Besides, we investigate the dependence of galaxy 
properties on the cosmic web environment, which is composed of knots, filaments,
sheets, and voids, classified by the Hessian matrix of the density 
field \citep{Zhang2009}.
The subhalo formation time is defined as the time at which the main branch 
progenitor reached half of its maximum mass. The galaxy stellar mass is estimated 
from \citet{Chang2015}, and the subhalo mass is obtained from the matched subhalo in
ELUCID simulation by the neighborhood abundance matching method \citep{Yang2018}.
Due to the distinct assembly history of centrals and satellites, we investigated the properties of centrals and satellites separately.
The following are our main findings:

\begin{itemize}

\item Central and satellite subhaloes have different formation time,
and satellite subhaloes are older than central subhaloes at fixed mass. Central 
and satellite galaxies have different stellar-to-subhalo mass ratios at fixed 
subhalo mass, especially for galaxies in low-mass subhaloes. At fixed subhalo
mass, satellite galaxies have significantly larger stellar mass than central
galaxies, especially for galaxies in subhaloes with mass less than $10^{12} \msun$.

\item Galaxy stellar-to-subhalo mass ratio is tightly correlated with the subhalo formation time, and the formation redshifts 
increase with stellar-to-subhalo mass ratios, especially for galaxies in high-mass subhaloes. 

\item Subhalo formation time is dependent on the cosmic web environment. 
For central subhaloes, there is a characteristic subhalo mass of $\sim 10^{12} \msun$, 
below which subhaloes in knots are older than subhaloes of the same mass in filaments, 
sheets or voids, while above which it reverses. For satellite subhaloes in low-mass 
range, subhaloes in knots are also oldest than those in other environments.

\item The cosmic web environmental dependence of galaxy stellar-to-subhalo mass
ratio is similar to that of the subhalo formation time. For centrals, there is
a characteristic subhalo mass of $\sim 10^{12} \msun$, below which the ratio 
decreases from knots to filaments to sheets to voids, above which it reverses.

\item For centrals or satellites in subhaloes with mass less than $10^{12} \msun$, 
galaxies in knots are redder than those in filaments, sheets, or voids. 
Above $10^{12} \msun$, there is no clear difference of galaxy colors in different 
cosmic web environment.

\item Satellite fraction is strongly dependent on the cosmic web environment, 
and decreases from knots to filaments to sheets to voids, especially for low-mass 
galaxies.

\end{itemize}

To conclude, we remark that galaxy stellar-to-subhalo mass ratio is a good proxy
of the subhalo formation time, especially for galaxies in high-mass subhaloes.
The environmental dependence of formation time is similar to that of galaxy 
stellar-to-subhalo mass ratio. Low-mass subhaloes show a decreasing formation 
redshifts and galaxy stellar-to-subhalo mass ratios from knots to voids, while the trend
reverses in high-mass subhaloes.

\section*{Acknowledgements}

We thank the anonymous referee for helpful comments that significantly 
improve the presentation of this paper.
This work is supported by the national natural science foundation of China 
(Nos. 11833005,  11890692, 11621303), 111 project No. B20019 and
Shanghai Natural Science Foundation, grant No. 15ZR1446700.  We acknowledge the science research grants from the China Manned Space Project with NO. CMS-CSST-2021-A02. 

This work is also supported by the High Performance Computing Resource
in the Core Facility for Advanced Research Computing at Shanghai
Astronomical Observatory.

Funding for the Sloan Digital Sky Survey IV has been provided by the
Alfred P. Sloan Foundation, the U.S. Department of Energy Office of
Science, and the Participating Institutions. SDSS acknowledges support
and resources from the Center for High-Performance Computing at the
University of Utah. The SDSS web site is www.sdss.org.

SDSS is managed by the Astrophysical Research Consortium for the
Participating Institutions of the SDSS Collaboration including the
Brazilian Participation Group, the Carnegie Institution for Science,
Carnegie Mellon University, the Chilean Participation Group, the
French Participation Group, Harvard-Smithsonian Center for
Astrophysics, Instituto de Astrof{\'i}sica de Canarias, The Johns
Hopkins University, Kavli Institute for the Physics and Mathematics of
the Universe (IPMU)/University of Tokyo, Lawrence Berkeley National
Laboratory, Leibniz Institut f{\"u}r Astrophysik Potsdam (AIP),
Max-Planck-Institut f{\"u}r Astronomie (MPIA Heidelberg),
Max-Planck-Institut f{\"u}r Astrophysik (MPA Garching),
Max-Planck-Institut f{\"u}r Extraterrestrische Physik (MPE), National
Astronomical Observatories of China, New Mexico State University, New
York University, University of Notre Dame, Observat{\'o}rio Nacional/
MCTI, The Ohio State University, Pennsylvania State University,
Shanghai Astronomical Observatory, United Kingdom Participation Group,
Universidad Nacional Aut{\'o}noma de M{\'e}xico, University of
Arizona, University of Colorado Boulder, University of Oxford,
University of Portsmouth, University of Utah, University of Virginia,
University of Washington, University of Wisconsin, Vanderbilt
University, and Yale University.

\section*{Data availability}
The data underlying this article will be shared on reasonable request to the corresponding author.

\bibliographystyle{mnras}
\bibliography{bibliography}

\bsp    
\label{lastpage}
\end{document}